\documentclass[prb,twocolumn,preprintnumbers]{revtex4-1}
\usepackage{amsmath,bm,xspace,SIunits,braket,color,graphicx,physics}

\pdfoutput=1

\begin{document}

\preprint{SAND2019-14414 J}

\title{Thermodynamics of the insulator-metal transition in dense liquid deuterium}
\date{\today}

\author{M.P. Desjarlais}
\email{mpdesja@sandia.gov}
\affiliation{Sandia National Laboratories, Albuquerque, New Mexico 87185, USA}

\author{M.D. Knudson}
\affiliation{Sandia National Laboratories, Albuquerque, New Mexico 87185, USA}

\author{R. Redmer}
\affiliation{Institute of Physics, University of Rostock, Rostock, Germany}






\begin{abstract}
Recent dynamic compression experiments [M. D. Knudson~\textit{et al.}, Science {\bf 348}, 1455 (2015); 
P. M. Celliers~\textit{et al.}, Science {\bf 361}, 677 (2018)]
have observed the insulator-metal transition in dense liquid deuterium, but with an
approximately 95 GPa difference in the quoted pressures for the transition at comparable estimated temperatures. 
It was claimed in the latter of these two papers that a very large latent heat effect on the temperature was overlooked 
in the first, requiring correction of those temperatures downward by a factor of two, thereby putting both experiments
on the same theoretical phase boundary and reconciling the pressure discrepancy.
We have performed extensive path-integral molecular dynamics calculations with density functional theory to directly calculate the isentropic temperature drop due to latent heat in the insulator-metal transition for dense liquid deuterium and show that this large temperature drop is not consistent with the underlying thermodynamics.
\end{abstract}

\maketitle 

\section{Introduction}

The long-standing quest to produce metallic hydrogen in the laboratory has progressed rapidly in the last few years. Recent improvements in static compression techniques, using diamond anvil cells (DAC) coupled with pulsed laser heating, have enabled investigation of the insulator-metal transition both in the cold, dense fluid\cite{Dzyabura, Ohta,Zaghoo,McWilliams,Zaghoo2} and in the solid\cite{Dias}. Of the various attempts to achieve metallization in hydrogen, two experimental approaches have used dynamic compression of liquid deuterium to produce multi-megabar pressures\cite{KnudsonSci,Celliers}. Similar in concept, these experiments were designed to probe the metallization of dense liquid deuterium by inducing a small shock followed by nominally isentropic compression. Through this combination of shock and ramp compression, the experiments, by design, avoid the solid phases of deuterium entirely while still maintaining a compression path well below the theoretical critical point temperature on the first-order insulator-metal phase transition boundary.

Experiments on the Sandia Z machine, as reported in Knudson \textit{et al.}~\cite{KnudsonSci} observed indications of a first-order insulator-metal transition in dense liquid deuterium at around 280 to 305 GPa. Subsequent dynamic compression experiments at the National Ignition Facility (NIF), reported in Celliers \textit{et al.}~\cite{Celliers} also found indications of an insulator-metal transition, but with a reported pressure of around 200 GPa. In the Z experiments, the pressure of the transition was deemed to be marked by the rapid drop in reflectivity upon release from a saturated high reflectivity state; it was argued that this provided the clearest signature of the transition, as transients in the system (particularly due to effects of thermal condition) would have ample time to damp out during the several tens of nanoseconds during which the system was in the metallic state. Conversely, in the NIF experiments the transition was denoted to correspond to a reflectivity value of 30\% during the initial rise in reflectivity of the compressed deuterium and did not provide data on the release from high pressure.
In this context, it is worth noting that the Z experiments indicate both a rise in reflectivity with increasing pressure as well as the abrupt fall with decreasing pressure. In principle, one can argue that the Z experiments also express the full extent of the phase boundary on the return back to low reflectivity.

Direct measurements of the temperature were not made in either of these experiments. Instead, the temperature was estimated using an equation of state (EOS) model for the shock ring-up, followed by estimates of the temperature increase for the quasi-isentropic ramp compression path using either (i) first-principles calculations (Z experiments), or (ii) an average of three tabulated EOS models (NIF experiments). We note that two of these three EOS models are PBE-based global models. It is well known that PBE\cite{Perdew} systematically underestimates the pressure conditions necessary for dissociation, and thus will predict isentropes that exhibit regions of strong $-dT/dP$ (due to latent heat of the transition) at pressures well below the actual metallization boundary; any global model that builds in latent heat well before the transition will underestimate the experimental temperature, perhaps by as much as several hundred K. With this caveat, the resulting estimates of the temperature on encountering the phase boundary were comparable for both sets of experiments, thus presenting $\sim$95 GPa difference in the location of the metallization boundary at comparable temperatures. 

In an effort to reconcile this significant difference in metallization pressure, Celliers \textit{et al.}~argue the NIF experiments probe the entrance of a given isentrope to the coexistence region while the Z experiments probe the exit from the coexistence region. This interpretation, as it stands, is consistent with the difference in the indicators used for identification of the phase boundary in the two sets of experiments: the rise of reflectivity with increasing pressure in the NIF experiments and the drop of reflectivity on the descent from high pressure in the Z experiments. However, it is further argued in Ref.~(\onlinecite{Celliers}) that a factor of two downward correction to the inferred temperatures reported by Knudson \textit{et al.}~\cite{KnudsonSci} is required due to a very large latent heat effect. This reanalysis, formally based on the Clausius-Clapeyron relation, but equivalent to requiring that the two well-separated pressure points ($\Delta P \sim 95$\ GPa) lie on the same phase boundary, would require that the temperature estimates in the Z experiments decrease by $\sim $600 K to nearly 900 K for the lowest and highest temperature loading paths, respectively. Subject to this reinterpretation both sets of experimental data would be in quite good agreement with recent coupled electron-ion Monte Carlo (CEIMC) calculations for the phase boundary reported by Pierleoni \textit{et al.}~\cite{Pierleoni}, and prior density functional calculations for the vdW-DF1 functional\cite{KnudsonSci}. While this result would be quite appealing from a theoretical point of view, we will show here that under more careful consideration this reinterpretation is in marked disagreement with the thermodynamics of the transition within a first-principles framework.

In particular, the Clausius-Clapeyron analysis presented in Ref.~(\onlinecite{Celliers}), by itself, is incomplete, as it does not include any constraints imposed by the thermodynamics of the phase transition. For a given phase boundary the pressure difference for the entrance and exit of an isentrope, the effective specific heat along the phase boundary, and the latent heat are not arbitrary, but rather dictated by thermodynamics. Here we demonstrate that while the various first-principles frameworks disagree on the location of the first-order transition boundary in $(P,T)$ space, they are in quite good agreement on the thermodynamics of the transition (i.e.~the specific heats, Gr\"uneisen gamma, bulk modulus, etc.). Furthermore, we show that subject to the thermodynamics of the transition, the large temperature drop suggested by Celliers \textit{et al.}~\cite{Celliers} is not thermodynamically consistent; such a large temperature drop would require an anomalously low specific heat for the metallic hydrogen phase. An initial critique of these temperature corrections on thermodynamic grounds was presented in Desjarlais \textit{et al.}~\cite{DesjarlaisComment}; counterarguments were presented in Celliers \textit{et al.}~\cite{CelliersResponse}. Here we provide a more detailed analysis, including nuclear quantum effects.

\begin{figure}
\centering
\label{fig:fig1}
\includegraphics[width=0.47 \textwidth]{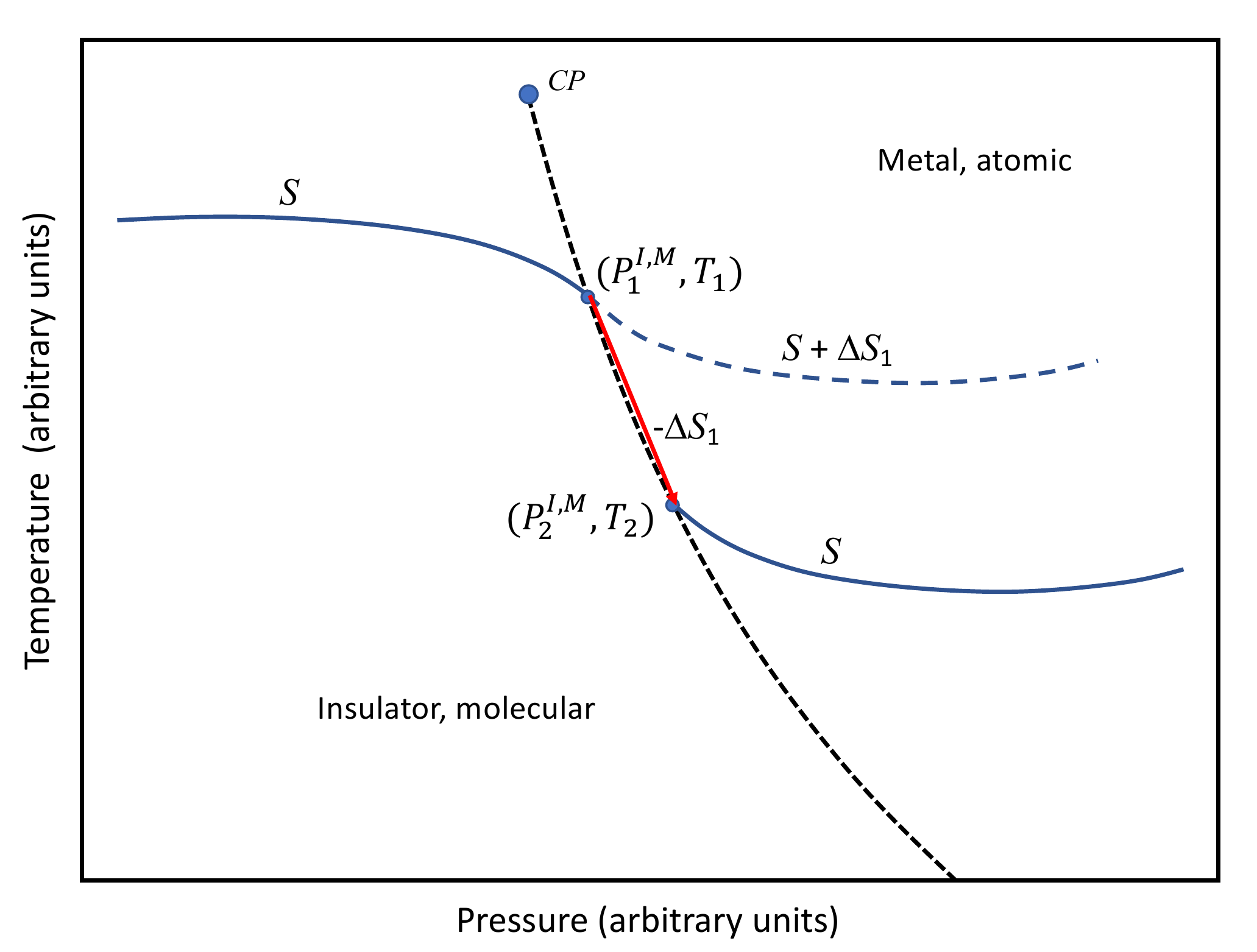}
\caption{Notional schematic of the $PT$ phase diagram for deuterium around the insulator-metal transition with representative isentropes. The black dashed line represents the phase boundary and terminates at the critical point. The isentropes are depicted with slight $-dT/dP$ near the phase boundary, resulting from a negative Gr\"uneisen $\gamma$, in accordance with first-principles calculations. The red arrow suggests a convenient integration path from $S + \Delta S_1$ back to $S$.}
\end{figure}

\section{Phase boundary thermodynamics}
To systematically address the question of the temperature drop resulting from the latent heat of the insulator-metal transition, we start with the assumption, common to both papers cited above, that the experimental path follows an isentrope.  Figure~1 is a schematic for the phase boundary and representative isentropes in $(P,T)$ space. As shown in Fig.~1, the isentropes exhibit a negative slope close to the transition, consistent with a negative Gr\"uneisen $\gamma $.  See, for example, the isentropes obtained by thermodynamic integration in Fig.~1 of Knudson \textit{et al.}~\cite{KnudsonSci}. For an isentrope $S$ that enters the coexistence region at a point $(P^I_1,T_1)$ and a given latent heat $\Delta H_1 = T_1 \Delta S_1$ we can readily compute the entropy at $(P^M_1,T_1)$ as $S + \Delta H_1/T_1 = S + \Delta S_1$.  Here the superscripts $I$ and $M$ refer to points on the insulator and metallic sides of the coexistence region respectively.  To obtain the temperature $T_2$ at which the isentrope $S$ exits the coexistence region at $(P^M_2,T_2)$ we need only compute the temperature drop required to remove the excess entropy $\Delta S_1$ in going from $(P^M_1,T_1)$ to $(P^M_2,T_2)$. Note that this path is a convenient integration path and should not be confused with the actual experimental thermodynamic path, as was done in Celliers \textit{et al.}~\cite{CelliersResponse}.

We derive in the following the exact expression for change in entropy along a line in $(P,T)$ space in terms of quantities readily
calculated within an $NVT$ first-principles framework.  We start with the total derivative of $S(P,T)$:
\begin{equation}
\label{eq:dS}
{dS} = \left({{\partial S}\over {\partial T}}\right)_{P} dT + \left({{\partial S}\over {\partial P}}\right)_{T} dP.
\end{equation}
We can specify an arbitrary line in $(P,T)$ space by imposing the linear constraint 
\begin{equation}
 {dP} = \left({dP \over dT}\right) _{line}  dT.
\end{equation}
This results in
\begin{equation}
\left({{dS}\over {dT}}\right)_{line} = {C_P\over T} + \left({{\partial S}\over {\partial P}}\right)_{T}\left({dP \over dT}\right)_{line}
\end{equation}
where we have used the definition
\begin{equation}
 \left({\partial S}\over {\partial T}\right)_{P} = {C_P\over T},
\end{equation}
where $C_P$ is the specific heat at constant pressure.
We can further reduce this in terms of quantities that are readily calculable with the Maxwell relation
\begin{equation}
\left({\partial S}\over {\partial P}\right)_{T} =  - \left({\partial V}\over {\partial T}\right)_{P} = \left({\partial P}\over {\partial T}\right)_{V}{\bigg/}\left({\partial P}\over {\partial V}\right)_{T}.
\end{equation}
Given the definitions
\begin{equation}
\label{gdef}
{{\gamma}}={V\over C_V}\left({\partial P \over \partial T}\right)_{V}=V\left({\partial P}\over {\partial E}\right)_{V}\ \ \   {\rm and} \ \ \  {{B_{T}\over V}}=-\left({\partial P}\over {\partial V}\right)_{T},
\end{equation}
where $B_T$ is the isothermal bulk modulus, $C_V$ is the specific heat at constant volume, and $\gamma$ is the Gr\"uneisen $\gamma$,
we arrive at
\begin{equation}
\label{eq:dSdT}
\left({dS}\over {dT}\right)_{line} = {C_P\over T}\Bigg[1 - {C_V\over C_P}{{\gamma T}\over B_T}  \left(d P \over d T\right)_{line}\Bigg].
\end{equation}
With calculations of $C_V$, $\gamma$, and $B_T$, $C_P$ is readily obtained via $C_P = C_V(1 + \gamma^2 T C_V/V B_T)$.
  
In Celliers \textit{et al.}~\cite{CelliersResponse} this expression (\ref{eq:dSdT}) was noted to be equivalent to Eq.~(4.19) in Reichl~\cite{Reichl}, 
but with the erroneous interpretation that this expression is only to be applied at constant volume. As is clear from the derivation, this expression for the change in entropy is completely general for a given line in $(P,T)$ space. For the specific application here,
this expression is applied along a line adjacent to, but on the metallic side of the phase transition, 
as indicated by the red arrow in Fig.~1.  For the slope of the coexistence boundary 
in the vicinity of 1300K to 1400 K, we use $dP/dT_{coex} =$ -0.12 GPa/K as an average, which agrees well with the 
published coexistence line for vdW-DF1, including nuclear quantum effects, around that temperature range~\cite{KnudsonSci}. $C_V$, $\gamma$, and $B_T$ are determined from first-principles calculations.

\section{First-principles calculations with nuclear quantum effects}
To obtain quantitative values for $C_V$, $\gamma$, and $B_T$, including nuclear quantum effects, we 
have performed several path integral molecular dynamics (PIMD) calculations with finite-temperature
 density functional theory as implemented in the VASP 5.3.5~\cite{Kresse1,Kresse2,Kresse3}, using the PIMD scheme of Alf{\`e} and Gillan~\cite{AlfePIMD}.  The DFT-PIMD calculations were performed with the vdW-DF1~\cite{DionDF1} 
exchange-correlation functional. Note the choice of vdW-DF1 for these calculations is motivated by arguments in 
Celliers \textit{et al.}~\cite{Celliers} that (\textit{i}) the onset of the phase transition is in close agreement with the phase boundary predicted by either CEIMC or DFT calculations with vdW-DF1 and nuclear quantum effects, and (\textit{ii}) that the Z experiments are indicative of the exit from that boundary.  Therefore vdW-DF1 is a logical choice for directly addressing the thermodynamics of their argument in that region of phase space in a density functional framework.  However, as discussed in Knudson \textit{et al.}~\cite{KnudsonSci} the phase boundary suggested by the Z experiments is in better agreement with the vdW-DF2 functional and the effect of the latent heat was implicitly computed in Knudson \textit{et al.}~\cite{KnudsonSci} through direct calculation of the isentropes on both sides of the phase boundary by thermodynamic integration.

Each DFT-PIMD calculation consists of 8 path integral molecular dynamics images for a Trotter time step $\leq 9.5 \times10^{-5}$ K$^{-1}$, with each image containing 256 deuterium atoms represented with a PAW potential~\cite{blochl, vasppaw}.
The Brillouin zone was sampled at the Baldereschi mean value point~\cite{Bald} and the plane wave cutoff energy was 700 eV.  
 \begin{figure} 
\centering
\label{fig:fig2}
\includegraphics[width=0.47 \textwidth]{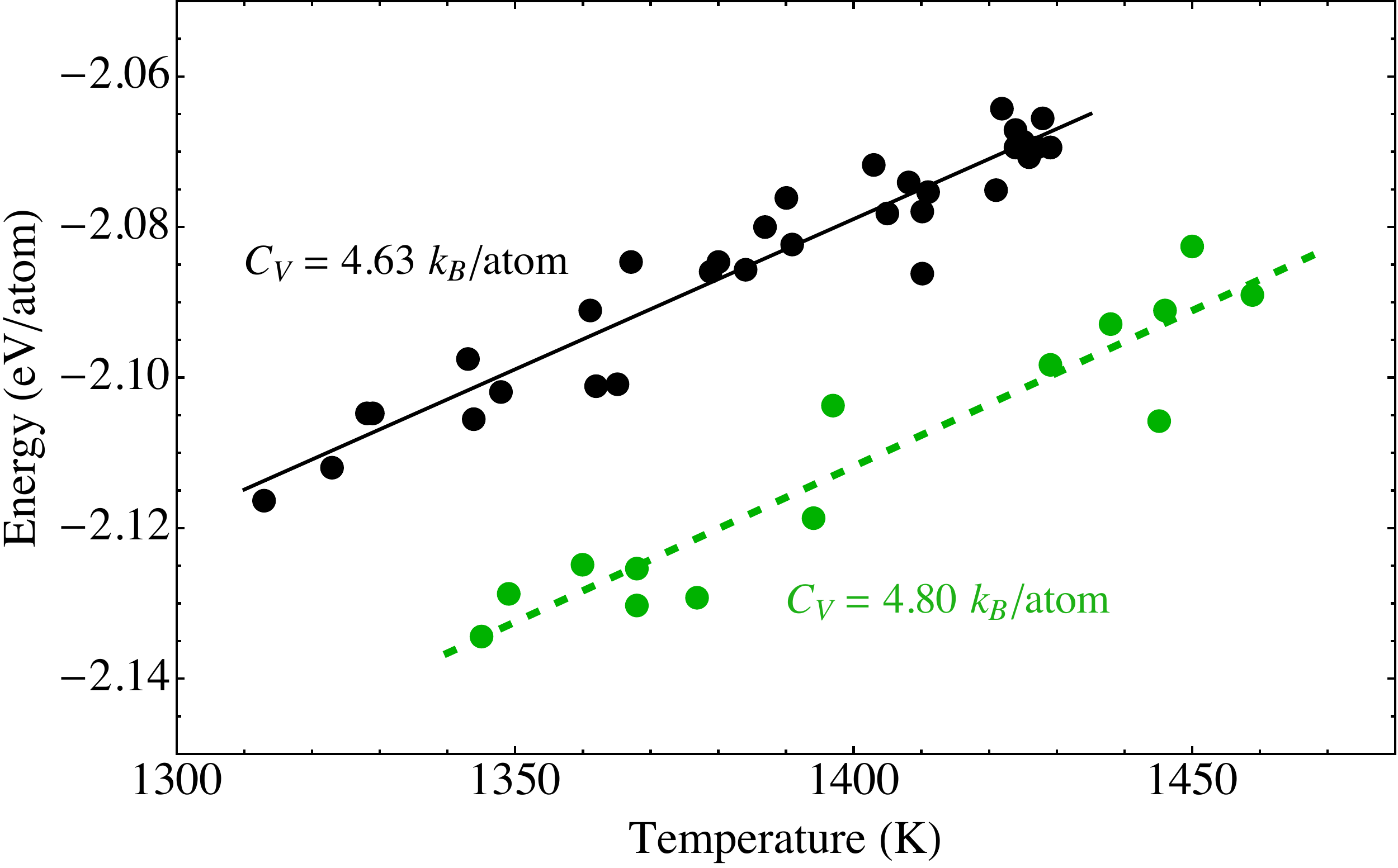}
\caption{Energy versus temperature in the vicinity of the insulator-metal transition at specific volumes of 1.80 (green circles, dotted fit) and 1.78 (black circles, solid fit) {\AA}$^3$/atom. $C_V$ is given by the slope.}
\end{figure}
Individual DFT-PIMD runs consisted of 4000 time steps of 0.25 fs, for a total simulated time of 1 ps.  An Andersen thermostat~\cite{Andersen} was employed to regulate the temperature and approximate a canonical ensemble.  Due to the relatively short nature of each individual run, 
 we observed statistical variation in the temperature about the target temperature, along with expected statistical variation in the thermodynamic quantities.  
 To minimize the statistical variance in the thermodynamic averages, generalized virial estimators~\cite{ParrinelloVirial} were used for the pressure and energy.  
 
 The results of these calculations for specific volumes of 1.80  {\AA}$^3$/atom (green circles, dotted fits) and 
 1.78  {\AA}$^3$/atom (black circles, solid fits) are illustrated in Fig.~2 as the total energy 
 versus temperature, and in Fig.~3 as pressure $\times$ volume versus the energy, providing $C_V$ and $\gamma$,
 respectively.  The lower temperature bound of the data set for each fit
 was chosen to correspond with a sharp drop in the dimer peak of the pair correlation function,
 as computed from the path centroid positions and indicative of the molecular to atomic transition.  
 By this construction we are performing calculations adjacent to the metallic side of the phase transition in the region traversed
 by the path from $(P_1^M, T_1)$ to $(P_2^M, T_2)$ suggested by the red arrow in Fig.~1. For a specific volume of 1.80 {\AA}$^3$/atom  
 we find a transition at $\sim$ 1345 K and $P$ = 204 GPa.  For a specific volume of 1.78 {\AA}$^3$/atom  
 we find a transition at $\sim$ 1310 K and $P$ = 208.5 GPa.  These $(P, T)$ phase transition points
 are in excellent agreement with the CEIMC predictions in Pierleoni \textit{et al.}~\cite{Pierleoni} for D$_2$ with quantum deuterons as well as the estimate from Knudson \textit{et al.}~\cite{KnudsonSci} for vdW-DF1 based on equating the spatial width of
the classical and quantum D$_2$ dimer peaks.
\begin{figure}
\centering
\label{fig:fig3}
\includegraphics[width=0.47 \textwidth]{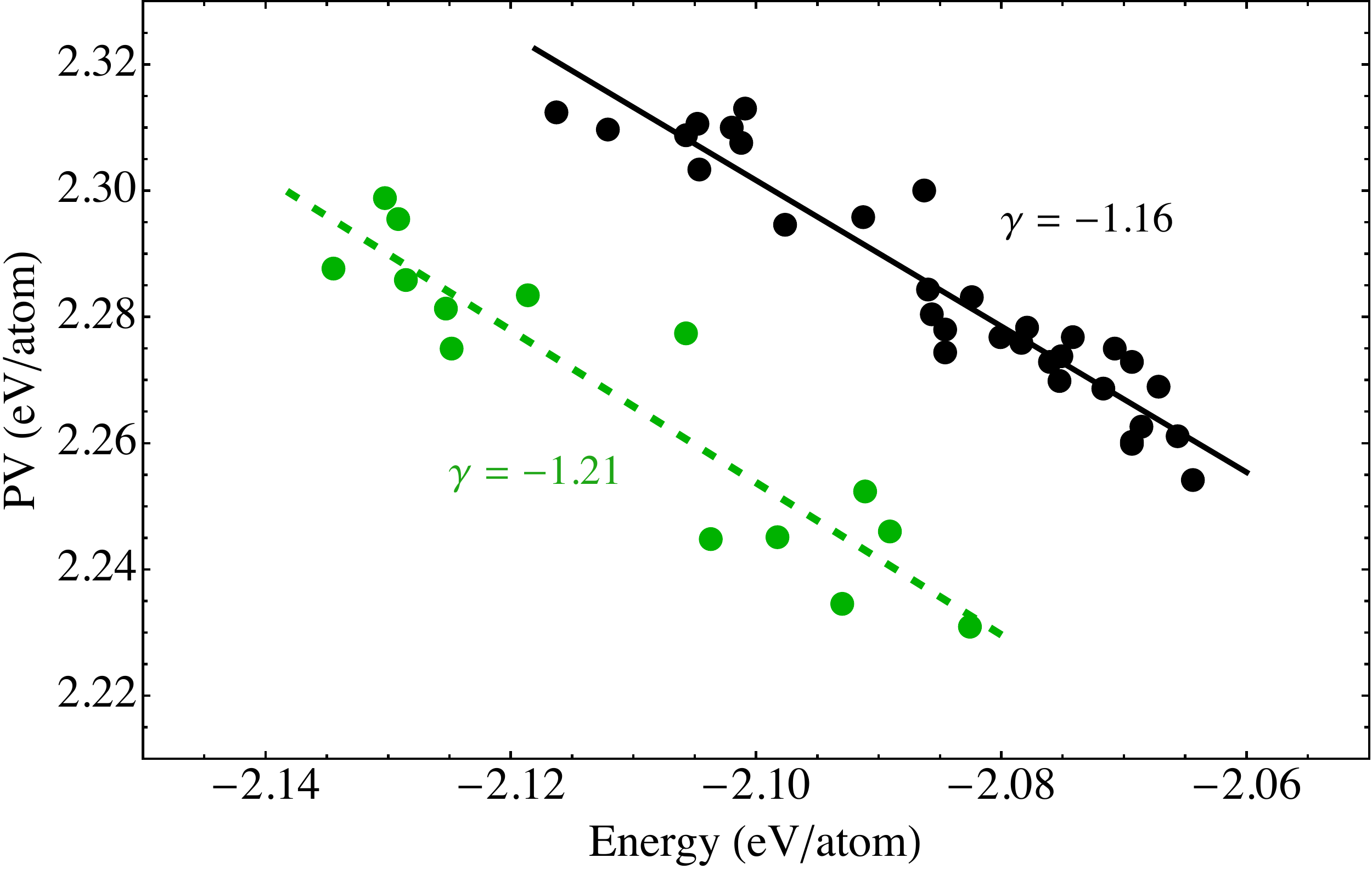}
\caption{Pressure $\times$ volume versus energy in the vicinity of the insulator-metal transition at specific volumes of 1.80 (green circles, dotted fit) and 1.78 (black circles, solid fit) {\AA}$^3$/atom. The Gr\"uneisen $\gamma $ is given by the slope.}
\end{figure}
 From the linear fits to the DFT-PIMD data
 we extract values of $C_V = 4.80\  { k_{B}/\rm atom}$ and $\gamma = -1.21$ for $V = 1.80\ ${\AA}$^3$/atom, and 
 $C_V = 4.63\  { k_{B}/\rm atom}$ and $\gamma = -1.16$ for $V = 1.78\ ${\AA}$^3$/atom.

 To determine $B_T$ adjacent to the phase boundary, we have computed $P$ versus $T$ over a wide range of volumes and temperatures as shown in Fig.~4.
 From the individual fits to the pressure, we generate $P$ versus $V$ at 1345 K, as shown by the corresponding 
 second order polynomial fit in Fig.~5. 
 Differencing the pressure fit at $V = 1.80\ ${\AA}$^3$/atom and using the second definition in Eq.~(\ref{gdef}) 
 yields $B_T = $ 272 GPa. 
 We find that $B_T$ increases rapidly over the course of several 10s of GPa, reaching 540 GPa within 50 GPa at 1345 K,
 however it is this lower local value that is relevant to the calculation along the phase boundary and which slightly depresses the effective specific heat along the coexistence line. 
\begin{figure}
\centering
\label{fig:fig4}
\includegraphics[width=0.47 \textwidth]{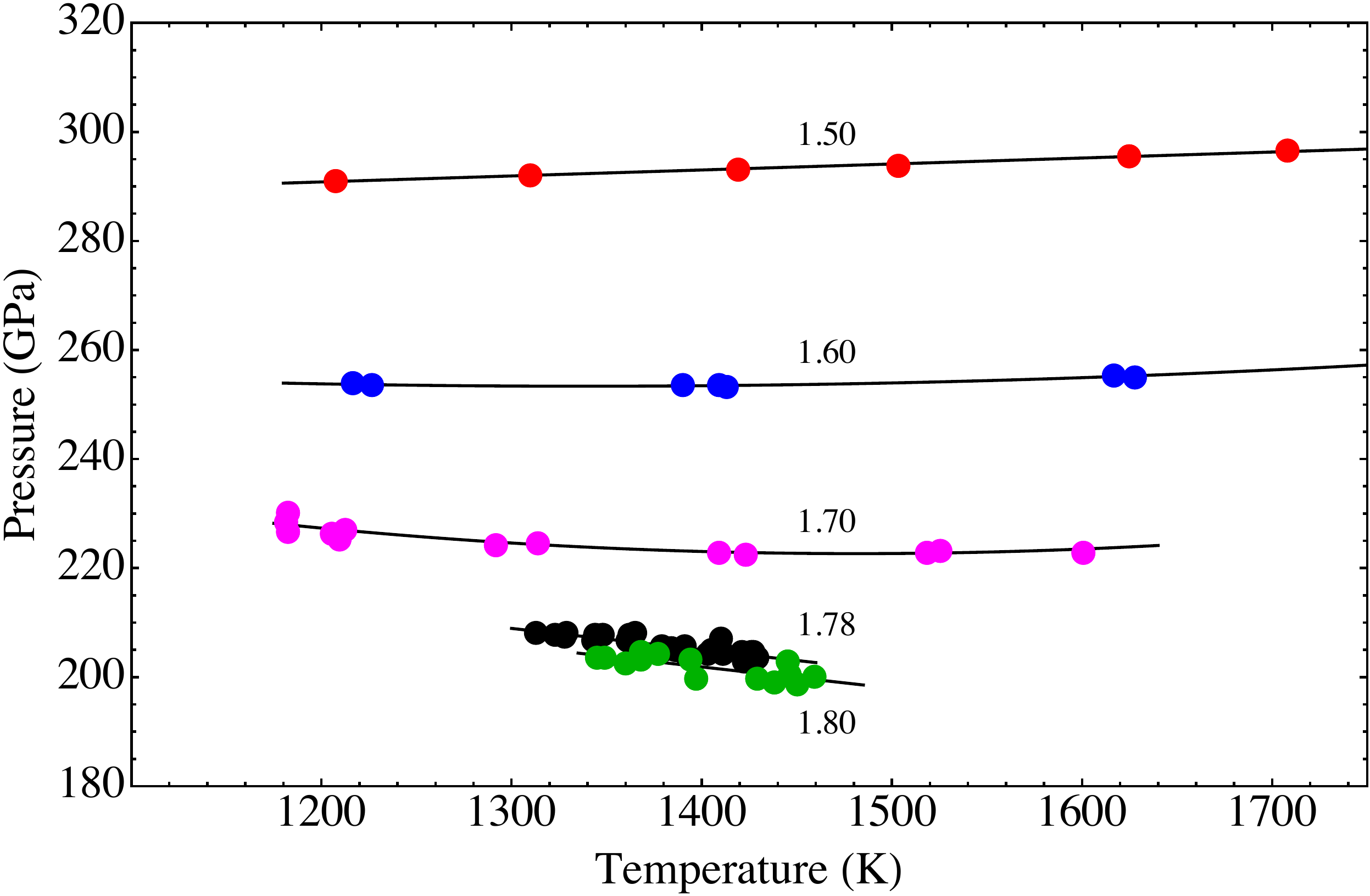}
\caption{Pressure versus temperature for several specific volumes. Labels for each fit indicate 
the specific volume in {\AA}$^3$/atom.}
\end{figure}

From $C_P = C_V(1 + \gamma^2 T C_V/V B_T)$, with $T= 1345$ K and $V=1.80\ {\AA}^3$/atom,  
we find $C_{P} = 6.07\  { k_{B}/\rm atom}$.
Combining terms and inserting in Eq.~(\ref{eq:dSdT}), we find
\begin{equation}
\label{eq:coex}
\left({dS}\over {dT}\right)_{coex} =  %
{C_P\over T}\Bigg[1 - {C_V\over C_P}{{\gamma T}\over B_T}  \left(d P \over d T\right)_{coex}\Bigg] \equiv {{C_{coex}}\over {T}},\end{equation}
with $C_{coex} = 2.64\  { k_{B}/\rm atom}$.  
Integrating Eq.~(\ref{eq:coex}) with the constraint $\int_{1}^{2} dS =- \Delta S_1$, and assuming $C_{coex}$ is weakly varying or 
represents an average value, gives
\begin{equation}
\label{eq:T2T1}
{{T_2}\over {T_1}} \cong \exp(-\Delta S_1/C_{coex}).
\end{equation}
 \begin{figure}
\centering
\label{fig:fig5}
\includegraphics[width=0.47 \textwidth]{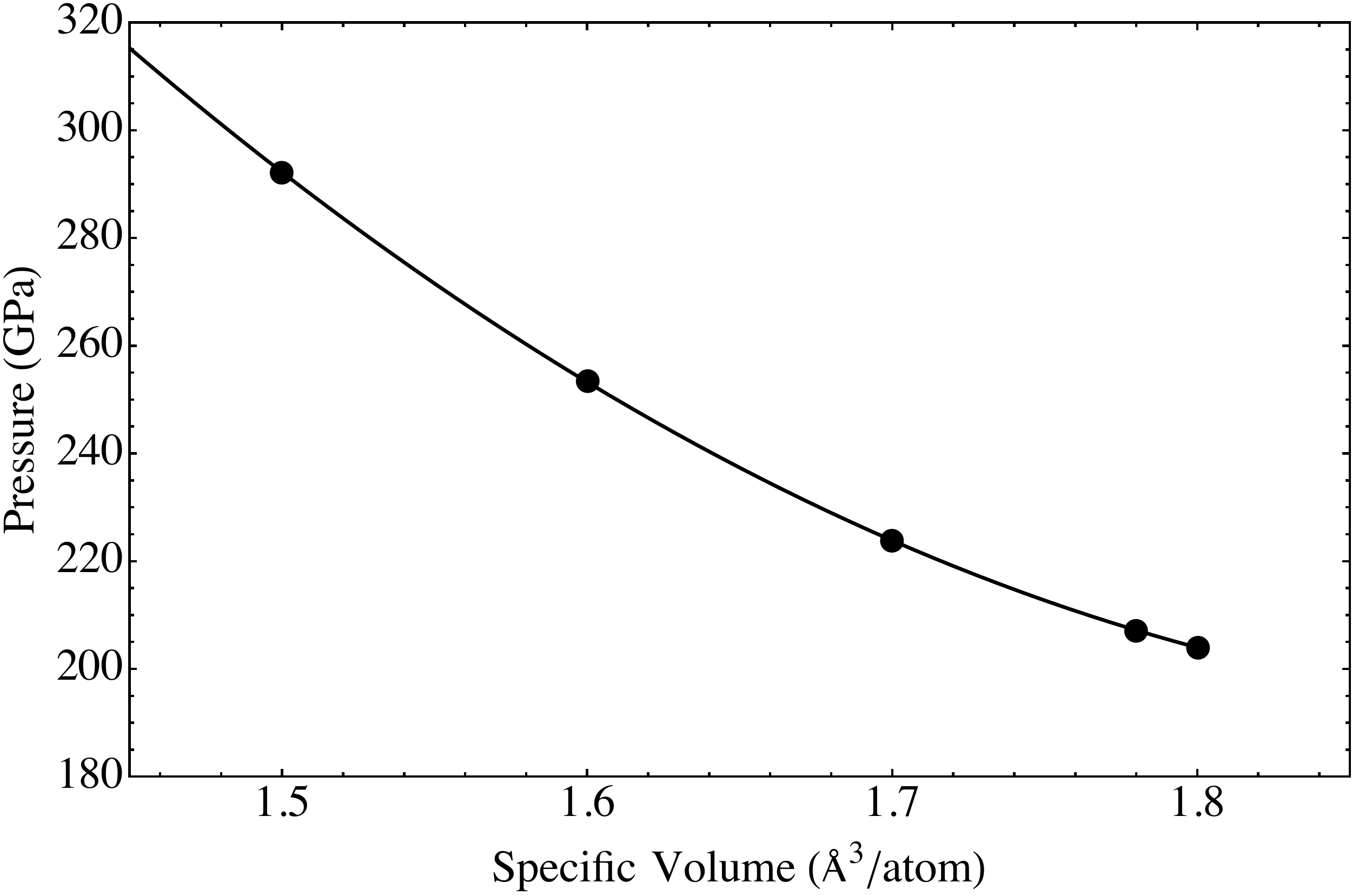}
\caption{Pressure versus specific volume at 1345 K.}
\end{figure}

\section{Isentropic temperature drop due to latent heat}
There are previous calculations of the latent heat for this insulator-metal transition in the literature~\cite{Morales, Pierleoni}.  We find,
consistent with the Pierleoni {\it et al.}~results for quantum protons, a latent heat of $\sim$0.05 $\pm $ 0.005 eV/atom (or 522~K to 
638~K), over the temperature range of 1300$-$1350~K.  Using $\Delta H_1=0.05$ eV/atom for the latent heat, and $T_1 =$ 1345 K,
sets $\Delta S_1 =$0.431 $k_{B}$/atom.  For $C_{coex}= 2.64\  {k_{B}/\rm atom}$ as calculated above at $V = 1.80\ ${\AA}$^3$/atom, 
${T_2}/{T_1}=0.85$, according to Eq.~(\ref{eq:T2T1}). We do not find these calculations very sensitive to the 
0.050 $\pm $ 0.005 eV/atom extrema quoted for the latent heats, producing only a $\mp$1.5\% change in ${T_2}/{T_1}.$

These final results are only very weakly dependent
on the explicit $T$ dependence within the square brackets of Eq.~(\ref{eq:coex}), so for the purposes of computations above we have treated $T$ as a constant equal to $T_1$.    
Because of the weak dependence, iteratively replacing $T$ by $\overline{T} = (T_1+T_2)/2$  
converges very rapidly.  As a specific quantitive example of this, 
consider a case where the calculated $C_{coex}= 2.64\  {k_{B}/\rm atom}$ at $V = 1.80\ ${\AA}$^3$/atom
is assumed to represent a midpoint or average value $\overline{C}_{coex}$ within the temperatures spanned from $T_1$ to $T_2$,  
where ${{T_2}/{T_1}} = \exp(-\Delta S_1/\overline{C}_{coex})$.
Iterating the above equations as described, with the target $\overline{T} = (T_1+T_2)/2 = 1345$ K, results in a solution with $T_1 = 1447$ K, 
$\Delta S_1 = $ 0.401 $k_{B}$/atom, $T_2 = 1243$ K,
and ${T_2}/{T_1}=0.86$. For a phase boundary slope of $dP/dT_{coex} =$ -0.12 GPa/K, this calculated $\Delta T =$ 202 K along the coexistence boundary would correspond to $\Delta P =$ 24 GPa. Comparing these results to the green curve in Fig.~1 of Knudson \textit{et al.}~\cite{KnudsonSci} 
reveals very good agreement with the latent heat induced temperature drop obtained there by thermodynamic integration
using the vdW-dF2 functional~\cite{LeeDF2}.  Furthermore, Figure~4c (green curve) of Knudson \textit{et al.}~\cite{KnudsonSci}, which shows reflectivity
as a function of pressure and corresponds to these same
approximate conditions, is in good agreement with this $\Delta P$ estimate on the fast rising linear portion of the reflectivity history.

It is interesting to note that despite the relatively high value of $C_{P} = 6.07\  { k_{B}/\rm atom}$ found on the metallic side
of the phase boundary, the negative
slope of the phase boundary and the negative slope of the isentrope near the boundary ($\gamma < 0$) 
combine to give an effective specific
heat through Eq.~(\ref{eq:T2T1}) that is very close to what we find well away from the phase boundary in the region where $\gamma \sim 0$.  In that region, at a specific volume of $1.60\ ${\AA}$^3$/atom and 1345 K (around 254 GPa in Fig.~4), the specific heat $C_{P} \approx C_V = 2.61\ {k_{B}/\rm atom}$.
As was noted in Desjarlais \textit{et al.}~\cite{DesjarlaisComment}, values of specific heat of this magnitude are expected for liquid alkali metals; $C_P=$ 3.5 $k_{B}$/atom for liquid lithium metal~\cite{Chase}.

\section{Discussion}
Given the close agreement between (\textit{i}) the results presented here using vdW-DF1, (\textit{ii}) the estimates provided in Desjarlais \textit{et al.}~\cite{DesjarlaisComment} obtained
using interpolations on CEIMC data from Pierleoni {\it et al.}~\cite{Pierleoni}, and (\textit{iii}) the direct isentrope calculations presented in Knudson \textit{et al.}~\cite{KnudsonSci} obtained through thermodynamic integration using vDW-DF2, we wish to
comment now on the apparent discrepancy between these results and the estimated temperature reduction factor ${T_2}/{T_1}=0.66$ presented by Celliers \textit{et al.}~\cite{CelliersResponse}. Their estimate was obtained from analysis of their Eqs.~(1) and (2). However, in computing their Eq.~(2), $\gamma$ was incorrectly treated as a constant over the entire 95 GPa range of the integral. 
As is clear from the temperature minimum of the isentropes obtained by thermodynamic integration in Fig.~1 of Knudson, \textit{et al.}~\cite{KnudsonSci} and suggested also by Fig.~S32 of the Supplementary Materials for Celliers \textit{et al.}~\cite{Celliers} along with
Fig.~1 of Celliers \textit{et al.}~\cite{CelliersResponse}, the Gr\"uneisen $\gamma \to 0$ and eventually turns positive 
over the course of several 10s of GPa beyond the phase boundary.  
This follows directly from $\left(\partial S/\partial P\right)_T = - \gamma C_V/B_T$ and is a direct consequence of exhausting the molecular to atomic transition. 
That $\gamma \to 0$ within approximately 50 GPa is illustrated in Fig.~4 where $\left(\partial P/\partial T\right)_V$ is essentially zero by 255 GPa
for temperatures between 1200 K and 1400 K. 
Correcting the integration in Eq.~(2) of Ref.~\onlinecite{CelliersResponse}, and accounting for the consequences in subsequent
steps in that analysis,
brings those estimates in line with the results obtained here.

A refinement of the finite-difference analysis outlined in Ref.~\onlinecite{Celliers} is presented in Ref.~\onlinecite{CelliersResponse}. This analysis
combines the Clausius-Clapeyron relation with the assumption that the NIF and Z
pressures are indicative of the entrance $(P^I_1,T_1)$ and exit $(P^M_2,T_2)$ of the coexistence region.  
By effectively constraining the 95 GPa pressure difference
to the coexistence line, they arrive at a temperature reduction factor due to latent heat of ${T_2}/{T_1}=0.58$.  This value of 0.58, when
combined with Eq.~(\ref{eq:T2T1}), and $T_1 = 1447$ K, implies an anomalously low $C_{coex}= 0.74\  {k_{B}/\rm atom}$, in gross disagreement with the $C_{coex}= 2.64\  {k_{B}/\rm atom}$ obtained from our first-principles calculations. We emphasize that this larger value for $C_{coex}$ is consistent across vdW-DF1, vdW-DF2, and CEIMC; even though these different first-principles frameworks disagree on the precise location of the insulator-metal phase transition boundary, they are in quite good agreement on the thermodynamics of the transition (i.e. the specific heats, Gr\"uneisen gamma, bulk modulus, etc.).

As noted in the Introduction, the criteria for identifying the location of the phase boundary are substantively different between the Z and NIF experiments. In the Z experiments the phase boundary was associated with the abrupt drop from a high reflectivity phase upon pressure release. The NIF experiments only probed the transition upon compression; there the phase boundary was associated with the reflectivity rising above a threshold of 30\%.  Both approaches are valid; however, comparing one to the other on unequal footing exacerbates the apparent discrepancy. Associating the phase boundary in the NIF experiments with a higher value of reflectivity, commensurate with completion of the transition, would result in a higher inferred value for the transition pressure. As a quantitative illustration, consider the calculations of Rillo {\it et al.}~\cite{Rillo}, which suggest a reflectivity in excess of 40\% at the completion of the transition at a temperature of 1500~K. Applying this criteria to the 1450~K (closest available and specifically N150914-2) reflectivity versus pressure trace in Fig.~2 of Celliers \textit{et al.}~\cite{Celliers} suggests a phase boundary exit pressure of 240 GPa.

Evaluating the completion of the transition would also result in a lower inferred temperature, in accordance with the latent heat considerations explored here. However, regarding the estimated temperature, as noted in the Introduction two of the three equation of state (EOS) models used to infer the temperature in the NIF experiments include PBE latent heat contributions at pressures below the observed transition pressure (in either the Z or NIF interpretations) and therefore result in lower predicted temperatures. The 2003 deuterium EOS of Kerley\cite{Kerley2003}, the same EOS used for estimating temperatures for the Z experiments prior to switching to DFT calculations of the isentrope, is much closer to, and slightly exceeds, the upper bound of the temperature estimates provided in Fig.~3 of Celliers \textit{et al.}~\cite{Celliers}. Calculation of the temperature path for this reported 1450 K transition case with the Kerley 2003 EOS is shown in Fig.~S18 of the Supplemental Material for Celliers \textit{et al.}~\cite{Celliers}; subtracting 200~K for latent heat at 240 GPa, as indicated by the calculations in this paper, suggests a temperature closer to 1625~K. We note that a transition at 240 GPa and 1625~K is much closer to the vdW-DF2 phase boundary than that of vdW-DF1. Interpreting this NIF experiment on the same footing as the Z experiments reduces the apparent discrepancy between the two experiments by about a factor of two, to approximately 42~GPa.  The remaining differences are not insignificant, but will require future experiments and analyses to reconcile.

\section{Conclusion}
We have performed an extensive study of the thermodynamics of the insulator-metal transition in dense liquid deuterium within a first-principles framework to assist in interpreting recent dynamic compression experiments. Specifically, we used density functional theory, including nuclear quantum effects, to directly calculate the temperature drop for an isentrope that traverses the first-order insulator-metal transition. This was accomplished by evaluating an exact expression for change in entropy along a line in $(P,T)$ space in terms of quantities readily calculated within an $NVT$ first-principles framework. An extensive set of path-integral molecular dynamics calculations with the vdW-DF1 functional were performed to obtain quantitative values for $C_V$, $\gamma$, and $B_T$ adjacent to the metallic side of the phase transition in the region traversed in recent dynamic compression experiments. The resulting temperature drops were found to be consistent with previous direct isentrope calculations~\cite{KnudsonSci} obtained through thermodynamic integration and estimates~\cite{DesjarlaisComment} based on interpolations of CEIMC data~\cite{Pierleoni}. Furthermore, these temperature drops are in stark disagreement with a recent reinterpretation presented by Celliers \textit{et al.}~\cite{Celliers,CelliersResponse}

The arguments presented in Celliers \textit{et al.}~\cite{Celliers,CelliersResponse} have their root in constraining both
the Z and NIF experiments to a given theoretical phase boundary and approximate isentrope, with the NIF experiments marking the entrance to the
coexistence region and the Z experiments probing the exit.  For the quoted pressures in the two
sets of experiments, forcing the Z experiments to the phase boundary would require a factor of two correction
downward in the temperature estimates for the Z experiments. However, as is clear from the analysis presented
here, any given pressure difference between the two experiments would, by this enforced temperature constraint,
result in a different effective specific heat along the phase boundary, generally in conflict with the
underlying thermodynamics.  The large temperature corrections required
for the estimated 95 GPa pressure difference would necessitate an anomalously low specific heat that is in gross
disagreement with first-principles calculations presented here using vdW-DF1, those in Pierleoni \textit{et al.}~\cite{Pierleoni} using CEIMC as argued in Desjarlais \textit{et al.}~\cite{DesjarlaisComment},
and our earlier direct calculations of the isentropes\cite{KnudsonSci} obtained through thermodynamic integration using vdW-DF2. While interpreting the 1450~K NIF experiment in a manner analogous to that used for the Z experiments suggests a substantially smaller discrepancy between the two sets of experiments, an approximately 42 GPa difference remains.  
What is clear from the analysis presented here is that the pressure difference cannot be explained through the supposition of a very large latent heat effect between the onset and completion of the phase transition.

\section*{Acknowledgments}
MPD would like to thank Dario Alf{\`e} for providing many of the path-integral algorithms.
Sandia National Laboratories is a multi-mission laboratory managed and operated by National Technology and Engineering Solutions of Sandia, LLC., a wholly owned subsidiary of Honeywell International, Inc., for the U.S. Department of Energy's National Nuclear Security Administration under Contract No.~${\rm DE}$-${\rm NA0003525}$. This paper describes objective technical results and analysis. Any subjective views or opinions that might be expressed in the paper do not necessarily represent the views of the U.S. Department of Energy or the United States Government. RR also thanks the Deutsche Forschungsgemeinschaft (DFG) for support via the SFB 652 and FOR 2440.

\newpage

\bibliographystyle{apsrev4-1}

\end{document}